\documentclass[pra, aps, twocolumn, floatfix, showpacs]{revtex4}
\usepackage{graphicx, amsmath, amssymb, times}

\begin{document}
\title{Vortex line in spin-orbit coupled atomic Fermi gases}
\author{M. Iskin}
\affiliation{
Department of Physics, Ko\c c University, Rumelifeneri Yolu, 34450 Sar{\i}yer, Istanbul, Turkey.
}
\date{\today}

\begin{abstract}
It has recently been shown that the spin-orbit coupling gives rise to topologically-nontrivial  
and thermodynamically-stable gapless superfluid phases when the pseudo-spin populations
of an atomic Fermi gas is imbalanced, with the possibility of featuring Majorana 
zero-energy quasiparticles. In this paper, we consider a Rashba-type spin-orbit 
coupling, and use the Bogoliubov-de Gennes formalism to analyze a single vortex line 
along a finite cylinder with a periodic boundary condition. We show that the signatures 
for the appearance of core- and edge-bound states can be directly found in the density
of single-particle states and particle-current density. In particular, we find that the 
pseudo-spin components counterflow near the edge of the 
cylinder, the strength of which increases with increasing spin-orbit coupling.
\end{abstract}

\maketitle

\section{Introduction}
\label{sec:intro}

Following the recent experimental success with artificial gauge fields and 
spin-orbit coupled atomic Bose gases~\cite{nist1, nist2}, there has been 
increasing theoretical interest in studying spin-orbit coupled atomic Fermi 
gases with balanced or imbalanced populations, at zero or finite temperatures,
in two or three dimensions, etc.~\cite{gong, shenoy, zhai, hui, subasi, wyi, 
carlos, he, ghosh, yang, zhou, duan, zhang, liao}.
The main motivation behind these works is that the spin-orbit coupled atomic 
Fermi gases are ideal systems for studying topologically-nontrivial superfluid 
phases~\cite{gong, subasi, wyi, carlos}, with the possibility of featuring 
Majorana zero-energy bound states for which the associated quasiparticle 
operators are self-Hermitian. This means that a zero-energy Majorana 
quasiparticle is its own anti-quasiparticle. Although these quasiparticles are predicted
to appear in low-dimensional strongly-correlated systems in various fields 
of physics, including the fractional quantum Hall systems~\cite{moore}, 
chiral two-dimensional superconductors~\cite{volovik, read}, chiral 
two-dimensional $p$-wave superfluids~\cite{tewari, mizushima}, 
three-dimensional topological insulator-superconductor heterostructures~\cite{fu}
one-dimensional nanowires~\cite{oreg, alicea}, 
spin-orbit coupled semiconductor-superconductor heterostructures~\cite{sau, mao}, etc., 
it has proved to be very difficult to realize them in these systems. 
Given that the cold atom systems offer unprecedented control in comparison 
to condensed matter ones, there is a good chance of creating and observing 
Majorana quasiparticles with atomic systems in the near future. 

\begin{figure} [htb]
\centerline{\scalebox{0.5}{\includegraphics{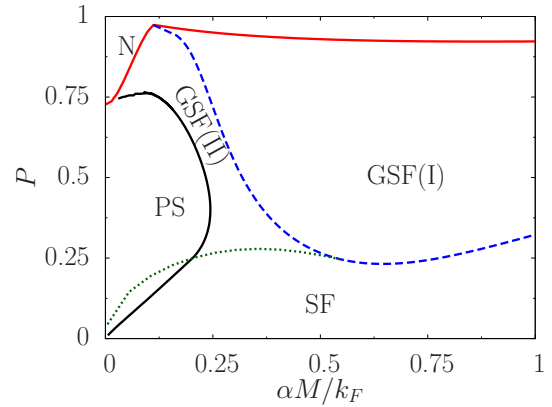}}}
\caption{\label{fig:pd} (Color online)
The mean-field ground-state phase diagram of a Rashba-type spin-orbit coupled 
Fermi gas is shown as a function of the population-imbalance parameter 
$P = (N_\uparrow - N_\downarrow)/N$ and the strength of the spin-orbit 
coupling $\alpha$ at unitarity, i.e. when $|a_s| \to \infty$. Here, N, SF, GSF, and 
PS correspond to normal, topologically-trivial gapped superfluid, 
topologically-nontrivial gapless superfluid, and phase separation, respectively.
(Adapted from Ref.~\cite{subasi}.)
}
\end{figure}

The first step in searching for the Majorana quasiparticles with spin-orbit coupled 
Fermi gases is to understand the phase diagram of these systems, which has recently 
been worked out within the mean-field approximation~\cite{gong, subasi, wyi, carlos}. 
For instance, the ground-state phase diagram of a Rashba-type spin-orbit 
coupling at unitarity, i.e. when the two-body scattering length $a_s$ between 
pseudo-spin components in vacuum diverges, is illustrated in Fig.~\ref{fig:pd}. 
There are three phases in the phase diagram~\cite{subasi, wyi, carlos}. 
While the normal (N) phase is characterized by a vanishing superfluid order parameter, 
the uniform superfluid and nonuniform superfluid, e.g. phase separation (PS), are 
distinguished by their thermodynamic stability when the order parameter is nonzero. 
Furthermore, in addition to the topologically-trivial gapped superfluid (SF) phase, 
the gapless superfluid (GSF) phase can be distinguished by the
momentum-space topology of its excitations. Depending on the number of 
zero-quasiparticle excitation energy regions in momentum space, there are two 
topologically-distinct gapless phases. For the Rashba-type spin-orbit coupling 
shown here, while GSF(II) has four zero-energy points, GSF(I) has only two.

The phase diagram illustrates that the spin-orbit coupling counteracts the population 
imbalance, and that this competition tends to stabilize the GSF phase against PS.
The anisotropic nature of the spin-orbit coupling (in momentum space) is also found 
to stabilize exotic superfluid phases. For instance, in sharp contrast to the $\alpha = 0$ 
case where only the gapless superfluid phase supports population imbalance, 
both the gapless and gapped superfluid phases are found to support population 
imbalance. Although Rashba-type spin-orbit coupling is considered 
in Fig.~\ref{fig:pd}, the topological structure shown here is quite robust against 
the effects of anisotropic spin-orbit couplings~\cite{subasi}. 

Since Majorana quasiparticles appear in the presence of topological defects, 
e.g. vortices, domain walls, boundaries between bulk phases, etc.,
one of the exciting experimental directions with spin-orbit coupled Fermi gases may 
be to create and observe a Majorana zero-energy quasiparticle bound to a vortex. 
For this purpose, in this paper, we consider a Rashba-type spin-orbit coupling, 
and use the Bogoliubov-de Gennes (BdG) formalism to analyze a single vortex 
line along a finite cylinder with a periodic boundary condition. 
We find signatures for the appearance of core- and 
edge-bound states in various observables, most notable of which is the counterflow 
of pseudo-spin particle-current densities near the edge of the cylinder.
The rest of the paper is organized as follows. First, we generalize the BdG formalism 
to spin-orbit coupled Fermi gases in Sec.~\ref{sec:bdg}, and then derive the 
self-consistency equations for a single vortex line in Sec.~\ref{sec:vortex}. 
We present the numerical solutions in Sec.~\ref{sec:numerics}, where we discuss 
the effects of spin-orbit coupling on the superfluid order parameter, particle 
density, energy spectrum, bound-state wave functions, single-particle density 
of states, and probability-current density. Our conclusions are briefly summarized 
in Sec.~\ref{sec:conc}.

\section{Bogoliubov-de Gennes Formalism}
\label{sec:bdg}

The results mentioned above are obtained by solving the self-consistent BdG 
equations, suitably generalized to spin-orbit coupled Fermi gases. Before 
presenting our numerical results, let us first present the theoretical framework 
of this formalism.

\subsection{Hamiltonian}
\label{sec:ham}
In this paper, we use the mean-field Hamiltonian density (in units of $\hbar = k_B = 1$),
$
H (\mathbf{r}) = \sum_{\sigma,\sigma'} \psi_\sigma^\dagger(\mathbf{r}) K_{\sigma\sigma'}(\mathbf{r}) \psi_{\sigma'}(\mathbf{r})
 + \Delta(\mathbf{r}) \psi_\uparrow^\dagger(\mathbf{r}) \psi_\downarrow^\dagger(\mathbf{r}) 
 + \Delta^*(\mathbf{r}) \psi_\downarrow(\mathbf{r}) \psi_\uparrow(\mathbf{r}),
$
describing two-component Fermi gases with attractive and short-range interactions.
Here, the operators $\psi_{\sigma}^\dagger (\mathbf{r})$ and $\psi_{\sigma} (\mathbf{r})$ 
create and annihilate a pseudo-spin $\sigma$ fermion at position $\mathbf{r}$, respectively, 
and $\Delta(\mathbf{r})$ is the mean-field superfluid order parameter. 
Furthermore, the operator
$
K_{\sigma\sigma}(\mathbf{r}) = -\nabla^2/(2M_\sigma) - \mu_\sigma
$
is the kinetic energy, where $M_\sigma$ is the mass and 
$\mu_\sigma$ is the chemical potential of $\sigma$ fermions, and the operator
$
K_{\uparrow\downarrow}(\mathbf{r}) = K_{\downarrow\uparrow}^\dagger(\mathbf{r}) = \alpha (p_y + ip_x)
$
is the spin-orbit coupling, where $\alpha \ge 0$ is its strength and $p_j = -i \partial/\partial j$
is the momentum operator.  Although we keep the formalism quite general, we present
numerical calculations only for the mass-balanced Fermi gases with $M_\uparrow = M_\downarrow = M$.

In the presence of a spin-orbit coupling, this Hamiltonian can be diagonalized via the 
generalized Bogoliubov-Valatin transformations,
$
\psi_\sigma(\mathbf{r}) = \sum_{n,\sigma'} [u_n^{\sigma\sigma'}(\mathbf{r}) \gamma_{n,\sigma'} + {v_n^{\sigma\sigma'}}^*(\mathbf{r}) \gamma_{n,\sigma'}^\dagger],
$
where $u_n^{\sigma\sigma'}(\mathbf{r})$ and $v_n^{\sigma\sigma'}(\mathbf{r})$ are the amplitudes,
and $\gamma_{n,\sigma}^\dagger$ and $\gamma_{n,\sigma}$ are the operators corresponding 
to the creation and annihilation of pseudo-spin $\sigma$ quasiparticles, respectively.
The resultant BdG equation can be written as
$
H (\mathbf{r}) \varphi_n^{\sigma'}(\mathbf{r}) = \varepsilon_n^{\sigma'} \varphi_n^{\sigma'}(\mathbf{r}),
$
where
\begin{equation}
\label{eqn:bdg}
H (\mathbf{r}) = 
\left[ \begin{array}{cccc}
K_{\uparrow\uparrow}(\mathbf{r}) & K_{\uparrow\downarrow}(\mathbf{r}) & 0 & \Delta(\mathbf{r}) \\
K_{\downarrow\uparrow}(\mathbf{r}) & K_{\downarrow\downarrow}(\mathbf{r}) & -\Delta(\mathbf{r}) & 0 \\ 
0 & -\Delta^*(\mathbf{r}) & -K_{\uparrow\uparrow}^*(\mathbf{r}) & -K_{\uparrow\downarrow}^*(\mathbf{r}) \\
\Delta^*(\mathbf{r}) & 0 & -K_{\downarrow\uparrow}^*(\mathbf{r}) & -K_{\downarrow\downarrow}^*(\mathbf{r})
\end{array} \right]
\end{equation}
is the Hamiltonian matrix given in the 
$
\varphi_n^{\sigma'}(\mathbf{r}) = [ u_n^{\uparrow\sigma'}(\mathbf{r}), u_n^{\downarrow\sigma'}(\mathbf{r}), 
v_n^{\uparrow\sigma'}(\mathbf{r}), v_n^{\downarrow\sigma'}(\mathbf{r}) ]^T
$
basis, and $\varepsilon_n^{\sigma'} \ge 0$ are the energy eigenvalues. Since the BdG equations do not 
depend on $\sigma'$, it is sufficient to solve only for $u_n^\sigma(\mathbf{r}) = u_n^{\sigma\sigma'}(\mathbf{r})$, 
$v_n^\sigma(\mathbf{r}) = v_n^{\sigma\sigma'}(\mathbf{r})$ and $\varepsilon_n = \varepsilon_n^{\sigma'}$.

\subsection{Self-consistency equations}
\label{sec:self}
Using the Bogoliubov-Valatin transformations, the mean-field superfluid order parameter
$
\Delta(\mathbf{r}) = g \langle \psi_\uparrow(\mathbf{r}) \psi_\downarrow(\mathbf{r}) \rangle,
$
where $g \ge 0$ is the strength of the attractive interaction between $\uparrow$ and $\downarrow$ 
fermions, and $\langle \cdots \rangle$ is the thermal average, becomes
$
\Delta(\mathbf{r}) = g \sum_n [u_n^\uparrow(\mathbf{r}) {v_n^\downarrow}^*(\mathbf{r}) f(\varepsilon_n)
+ u_n^\downarrow(\mathbf{r}) {v_n^\uparrow}^*(\mathbf{r}) f(-\varepsilon_n).
$
Here, $f(x) = 1/(e^{x/T} + 1)$ is the Fermi function and $T$ is the temperature. 
As usual, we relate the interaction strength $g$ to the two-body scattering length $a_s$ between 
an $\uparrow$ and a $\downarrow$ fermion  in vacuum via the relation,
$
1/g = -M_rV/(4\pi a_s) + \sum_{\mathbf{k}}1/(\varepsilon_{\mathbf{k},\uparrow} + \varepsilon_{\mathbf{k},\downarrow}),
$
where $M_r = 2M_\uparrow M_\downarrow/(M_\uparrow + M_\downarrow)$ is twice the 
reduced mass of an $\uparrow$ and a $\downarrow$ fermion, $V$ is the volume of the sample 
and 
$
\varepsilon_{\mathbf{k},\sigma} = k^2/(2M_\sigma)
$
is the kinetic energy. This leads to
$
g = 4\pi^2a_s / [ 2M_r a_s \sqrt{2M_r \varepsilon_c} - M_r \pi],
$
where $\varepsilon_c$ is the energy cutoff used in the $\mathbf{k}$-space integration 
(to be specified below in Sec.~\ref{sec:numerics}). 

To determine $\mu_\sigma$, the order parameter equation has to be solved self-consistently 
with the number equations
$
N_\sigma = \int d\mathbf{r} n_\sigma(\mathbf{r}),
$
where
$
n_\sigma(\mathbf{r}) = \langle \psi_\sigma^\dagger(\mathbf{r}) \psi_\sigma(\mathbf{r}) \rangle
$
is the local density of $\sigma$ fermions. Using the Bogoliubov-Valatin transformations,
we obtain
$
n_\sigma(\mathbf{r}) = \sum_n [ |u_n^\sigma(\mathbf{r})|^2 f(\varepsilon_n) 
+ |v_n^\sigma(\mathbf{r})|^2 f(-\varepsilon_n)].
$
Having generalized the BdG formalism to spin-orbit coupled Fermi gases, next 
we apply it for a single vortex line.

\section{Single vortex line}
\label{sec:vortex}
In particular, we consider a single vortex line positioned along a finite cylinder 
of radius $R$ and length $L$, and with a periodic boundary condition in the $z$ 
direction, in such a way that the order parameter can be written as
$
\Delta(\mathbf{r}) = \Delta(r) e^{-i\theta},
$
where $r$ and $\theta$ are the cylindrical coordinates $\mathbf{r} = (r, \theta, z)$~\cite{gygi}.
Note in this coordinate system that the spin-orbit coupling term becomes
$
K_{\uparrow \downarrow} (\mathbf{r}) = e^{-i\theta}[\partial/\partial r - (i/r)\partial/\partial \theta],
$
showing that the single vortex line has rotational invariance around the $z$ axis, 
so that the solutions of the BdG equation have a well-defined planar angular 
momentum $m$, i.e. $m$ is a good quantum number. In addition, the system is 
assumed to have translational invariance along the $z$ direction, i.e. $p_z$ 
momentum is also a good quantum number.

Thus, for a singly-quantized vortex line considered in this paper, we may choose the 
normalized wave functions as
$
u_n^\uparrow(\mathbf{r}) = u_{nms}^\uparrow(r) e^{im\theta} e^{ik_s z} /\sqrt{2\pi L}
$
and
$
v_n^\uparrow(\mathbf{r}) = v_{nms}^\uparrow(r) e^{i(m+2)\theta} e^{ik_s z}/\sqrt{2\pi L}
$
for the $\uparrow$ components, and
$
u_n^\downarrow(\mathbf{r}) = u_{nms}^\downarrow(r) e^{i(m+1)\theta} e^{ik_s z}/\sqrt{2\pi L}
$
and
$
v_n^\downarrow(\mathbf{r}) = v_{nms}^\downarrow(r) e^{i(m+1)\theta} e^{ik_s z}/\sqrt{2\pi L}
$
for the $\downarrow$ ones. Here, $k_s = 2\pi s/L$ is the wave vector along the $z$ direction
with $s = 0, \pm 1, \pm 2, \dots$. This particular choice (which is not unique~\cite{sau, mao}) 
allow us to decouple the BdG equations into independent subspaces of $(m, s)$ sectors. 
We further project the radial wave functions onto a set of Bessel functions normalized in 
a disc of radius $R$~\cite{gygi}, i.e.
$
\phi_{j,m}(r) = \sqrt{2} J_m(\beta_{j,m} r/R) / [R J_{m+1}(\beta_{j,m})],
$
where $j = 1, 2, 3, \dots$ and the argument $\beta_{j,m}$ is the $j$th zero of $J_m(x)$. 
More explicitly, we have
$
u_{nms}^\uparrow(r) = \sum_j c_{nmsj}^\uparrow \phi_{j,m}(r)
$
and
$
v_{nms}^\uparrow(r) = \sum_j d_{nmsj}^\uparrow \phi_{j,m+2}(r)
$
for the $\uparrow$ radial wave functions, and
$
u_{nms}^\downarrow(r) = \sum_j c_{nmsj}^\downarrow \phi_{j,m+1}(r)
$
and
$
v_{nms}^\downarrow(r) = \sum_j d_{nmsj}^\downarrow \phi_{j,m+1}(r)
$
for the $\downarrow$ ones, and they already satisfy the boundary conditions
$u_{nms}^\sigma(R) = v_{nms}^\sigma(R) = 0$ at the edge of the cylinder.

Using the orthonormality condition
$
\int_0^R r dr \phi_{j,m}(r) \phi_{j',m}(r) = \delta_{jj'}
$
where $\delta_{jj'}$ is the Kronecker delta, this procedure reduces the BdG equation 
given in Eq.~(\ref{eqn:bdg}) to a $4j_{max} \times 4j_{max}$ matrix eigenvalue problem,
\begin{align}
\label{eqn:bdg.matrix}
\sum_{j'} \left( \begin{array}{cccc}
K_{\uparrow,ms}^{jj'} & S_m^{jj'} & 0 & \Delta_m^{jj'} \\
S_m^{jj'} & K_{\downarrow,ms}^{jj'} & -\Delta_{m+1}^{jj'} & 0 \\ 
0 & -\Delta_{m+1}^{jj'} & -K_{\uparrow,ms}^{jj'} & S_{m+1}^{j'j} \\ 
\Delta_{m}^{jj'}& 0 & S_{m+1}^{j'j} & -K_{\downarrow,ms}^{jj'} 
\end{array} \right)
&
\left( \begin{array}{c}
c_{nmsj'}^\uparrow \\
c_{nmsj'}^\downarrow \\
d_{nmsj'}^\uparrow \\
d_{nmsj'}^\downarrow
\end{array} \right) 
\nonumber \\
= \varepsilon_{nms}
\left( \begin{array}{c}
c_{nmsj}^\uparrow \\
c_{nmsj}^\downarrow \\
d_{nmsj}^\uparrow \\
d_{nmsj}^\downarrow
\end{array} \right)
&,
\end{align}
for each $(m, s)$ sector, if we allow $1 \le j \le j_{max}$ states. Here,
$
K_{\sigma,ms}^{jj'} = [\beta_{j,m}^2/(2M_\sigma R^2) + k_s^2/(2M_\sigma) - \mu_\sigma] \delta_{jj'}
$
are the kinetic energy terms, 
$
S_m^{jj'} = \alpha \int_0^R r dr \phi_{j,m}(r) [\partial/\partial r + (m+1)/r] \phi_{j',m+1}(r)
$
are the spin-orbit coupling terms leading to
$
S_m^{jj'}= \alpha C_{j'm} \int_0^R r dr \phi_{j,m}(r) J_m(\beta_{j',m+1}r/R)
$
where 
$
C_{j'm} = \sqrt{2} \beta_{j',m+1} / [R^2 J_{m+2}(\beta_{j',m+1})],
$
and
$
\Delta_m^{jj'} = \int_0^R r dr \Delta(r) \phi_{j,m}(r) \phi_{j',m+1}(r)
$
are the pairing terms. The same procedure also reduces the order-parameter equation to
\begin{align}
\label{eqn:op}
&\Delta(r) = \frac{g}{2\pi L} \sum_{nmsjj'} [ c_{nmsj}^\downarrow d_{nmsj'}^\uparrow \phi_{j,m+1}(r) \phi_{j',m+2}(r) \nonumber \\
&\times f(\varepsilon_{nms}) + c_{nmsj}^\uparrow d_{nmsj'}^\downarrow \phi_{j,m}(r) \phi_{j',m+1}(r) f(-\varepsilon_{nms}) ],
\end{align}
and the local-density equations to
\begin{align}
\label{eqn:ne.up}
&n_\uparrow(r) = \frac{1}{2\pi L} \sum_{nmsjj'} [ c_{nmsj}^\uparrow c_{nmsj'}^\uparrow \phi_{j,m}(r) \phi_{j',m}(r) f(\varepsilon_{nms}) \nonumber \\
& + d_{nmsj}^\uparrow d_{nmsj'}^\uparrow \phi_{j,m+1}(r) \phi_{j',m+2}(r) f(-\varepsilon_{nms}) ], \\
\label{eqn:ne.down}
&n_\downarrow(r) = \frac{1}{2\pi L} \sum_{nmsjj'} [ c_{nmsj}^\downarrow c_{nmsj'}^\downarrow \phi_{j,m+1}(r) \phi_{j',m+1}(r) \nonumber \\ 
&\times f(\varepsilon_{nms}) + d_{nmsj}^\downarrow d_{nmsj'}^\downarrow \phi_{j,m+1}(r) \phi_{j',m+1}(r) f(-\varepsilon_{nms}) ].
\end{align}
We recall that the sums are only over the quasiparticle states with $\varepsilon_{nms} \ge 0$.
Using the orthonormality condition, we also obtain the total number of $\sigma$ fermions as
$
N_\sigma = \sum_{nmsj} [(c_{nmsj}^\sigma)^2 f(\varepsilon_{nms}) + (d_{nmsj}^\sigma)^2 f(-\varepsilon_{nms}) ].
$
We emphasize that these mean-field equations can be used for all values of $a_s$ 
and $\alpha$ at low $T$, but they provide only a qualitative description of the 
system outside of the weak-coupling regime, i.e. in the BCS-BEC crossover. 
In this paper, we set the temperature to zero, and consider a strongly-interacting 
Fermi gas at unitarity, i.e. $|a_s| \to \infty$, as a function of $\alpha$.

\section{Numerical Results}
\label{sec:numerics}

In our numerical calculations, we set a large energy cutoff $\varepsilon_c = 10\varepsilon_F$, and 
numerically solve the self-consistency Eqs.~(\ref{eqn:bdg.matrix})-(\ref{eqn:ne.down}) at $T = 0$.
Here, $\varepsilon_F = k_F^2/(2M)$ is a characteristic Fermi-energy scale where 
$k_F$ is the Fermi momentum corresponding to the bulk value of the total density of 
fermions, i.e. $n_\uparrow(r) + n_\downarrow(r) = k_F^3/(3\pi^2)$ at the bulk.
We also choose $R = 25/k_F$ as the radius and $L = 10/k_F$ as the length of the cylinder, 
and $j_{max} = 50$ and $|m|_{max} = 75$ as the maximum quantum numbers. Note that
$|s|_{max} = L\sqrt{M\varepsilon_c/2}/\pi$ in order to be consistent with the energy cutoff. 
Since the presence of a single vortex line can not significantly effect the bulk 
parameters, we first solve $\mu_\sigma$ and $\Delta_0$ self-consistently for a 
vortex-free thermodynamic system, and then use these solutions as an input for 
our vortex-line calculation,  where $\Delta_0$ corresponds to the bulk value of $\Delta(r)$. 
Here, we assume $\Delta(r)$ is real without losing generality.

\subsection{Order parameter and density of fermions}
\label{sec:density}

\begin{figure} [htb]
\centerline{\scalebox{0.6}{\includegraphics{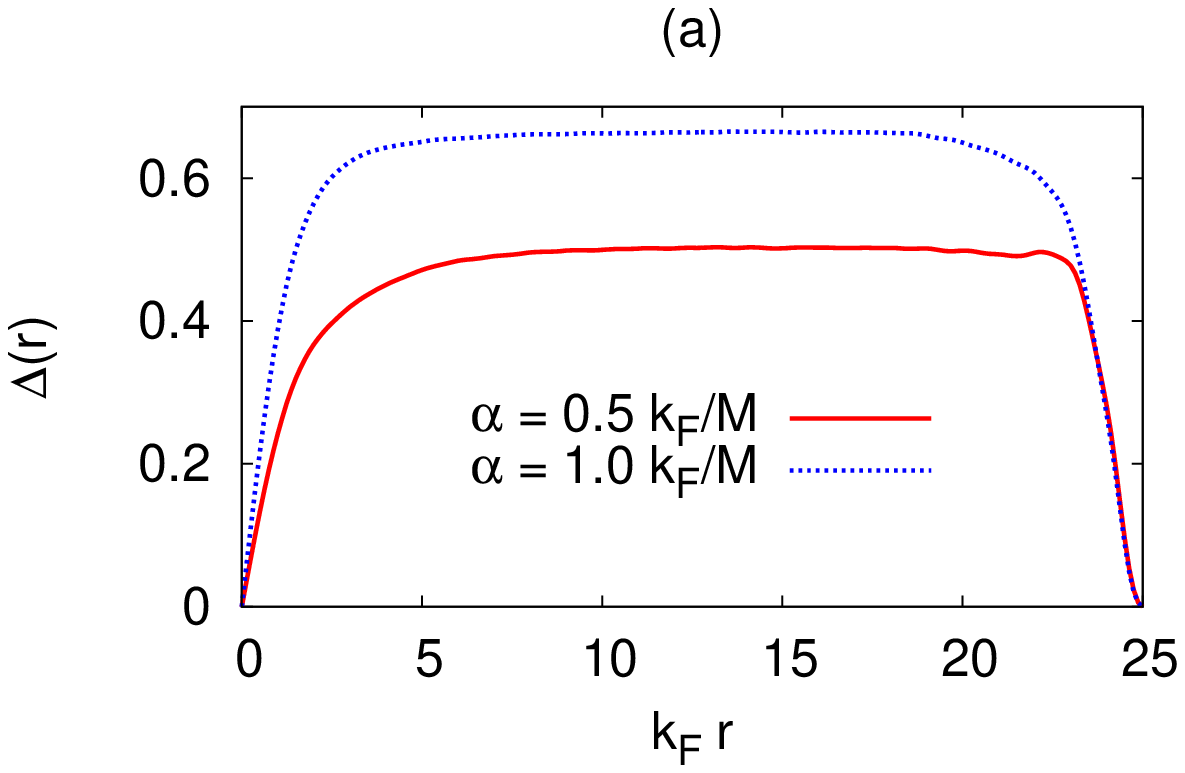}}}
\centerline{\scalebox{0.6}{\includegraphics{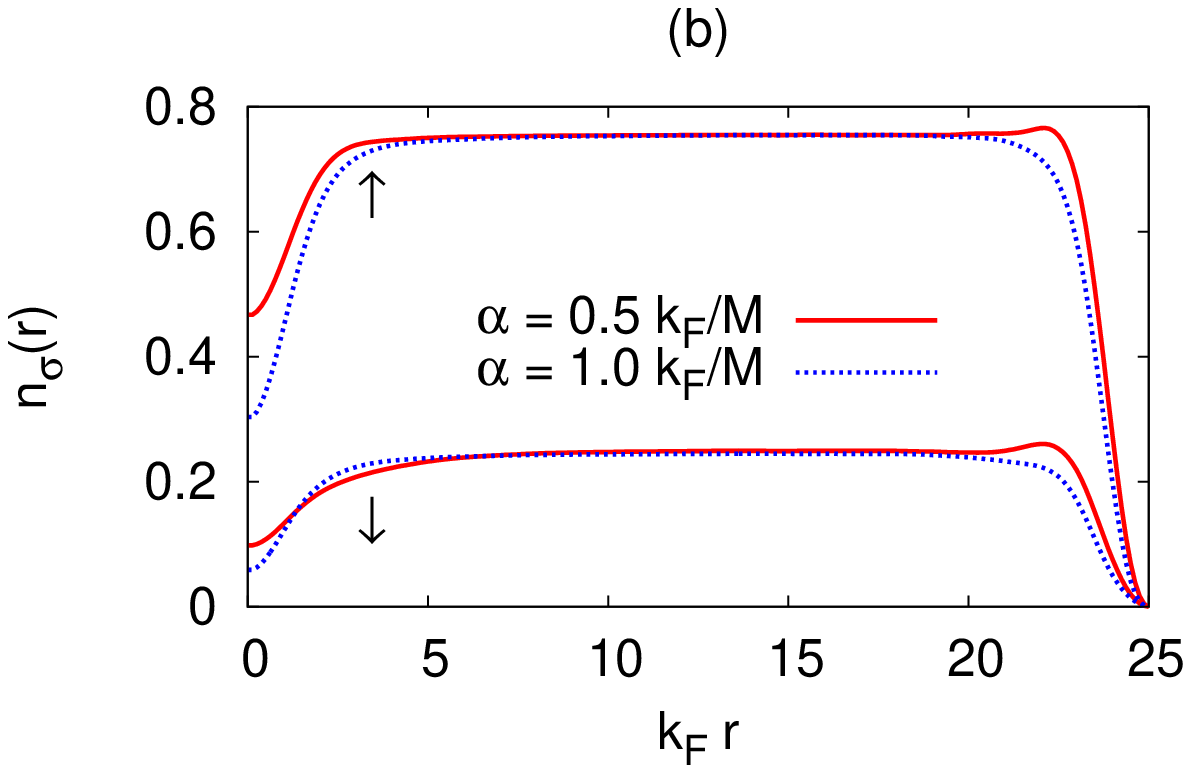}}}
\caption{\label{fig:density} (Color online) 
The order parameter $\Delta(r)$ (in units of $\varepsilon_F$) and density $n_\sigma(r)$ 
(in units of $k_F^3/(3\pi^2)$) profiles are shown as a function of radial distance $r$ 
(in units of $1/k_F$). Here, we set $1/(k_F a_s) = 0$ and $P = 0.5$, and vary $\alpha$.
}
\end{figure}

In Fig.~\ref{fig:density}(a), we show typical order-parameter profiles $\Delta(r)$ for 
$\alpha = 0.5k_F/M$ and $\alpha =k_F/M$ when $P = 0.5$, i.e. it rapidly increases from 
zero around the vortex core, saturates to its bulk value $\Delta_0$ around $k_F r \simeq 5$ and 
then it rapidly decreases to zero near the edge of the cylinder~\cite{gygi}. 
Here, the population-imbalance parameter 
$
P = [n_\uparrow(r) - n_\downarrow(r)] / [n_\uparrow(r) + n_\downarrow(r)]
$ 
is defined at the bulk.
We see that $\Delta(r)$ increases with increasing $\alpha$, e.g. its bulk value increases 
from $0.50\varepsilon_F$ to $0.66\varepsilon_F$, and that the effect of spin-orbit coupling is 
similar to the effect of increased interaction strength. This is due to the increased density 
of states with increasing $\alpha$, and it is consistent with the previous results on 
thermodynamic systems~\cite{gong, shenoy, zhai, hui, subasi, wyi, 
carlos, he, ghosh, yang, zhou, duan, zhang, liao}. 

In Fig.~\ref{fig:density}(b), we show the corresponding density profiles $n_\sigma(r)$ for 
the same parameters. When $\alpha = 0$, it is well-known for the fermion superfluids that 
the density is depleted near the vortex core compared to its bulk value~\cite{gygi}, 
and that the depletion increases with increasing interaction strength toward the 
molecular BEC side. 
This is because the energy separation between the vortex core-bound states
increases with increasing $\Delta_0$ which makes them less occupied.
In fact, for a population-balanced Fermi gas, the density depletes fully and becomes zero 
at the vortex core in the molecular BEC limit, consistent with the theory of weakly 
interacting atomic Bose gases. However, for population-imbalanced Fermi gases, the 
vortex core may still be filled with excess fermions toward this limit~\cite{takahashi}. When $\alpha \ne 0$, 
in Fig.~\ref{fig:density}(b), we again see that the effect of spin-orbit coupling is similar to the 
effect of increased interaction strength, i.e. density depletion also increases for both 
$\sigma$ components with increasing $\alpha$. To further understand the density 
depletions, next we analyze the spectrum of energy eigenvalues .

\subsection{Energy spectrum}
\label{sec:spectrum}

\begin{figure} [htb]
\centerline{\scalebox{0.6}{\includegraphics{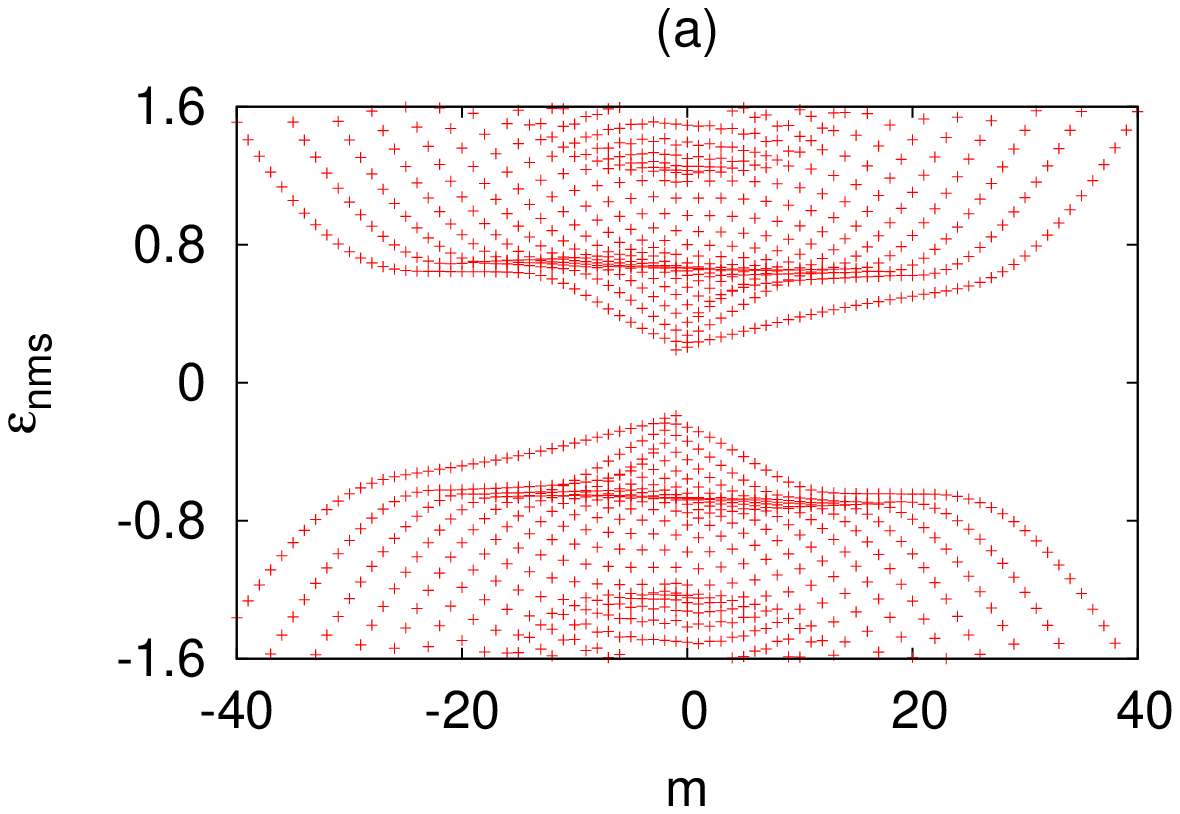}}}
\centerline{\scalebox{0.6}{\includegraphics{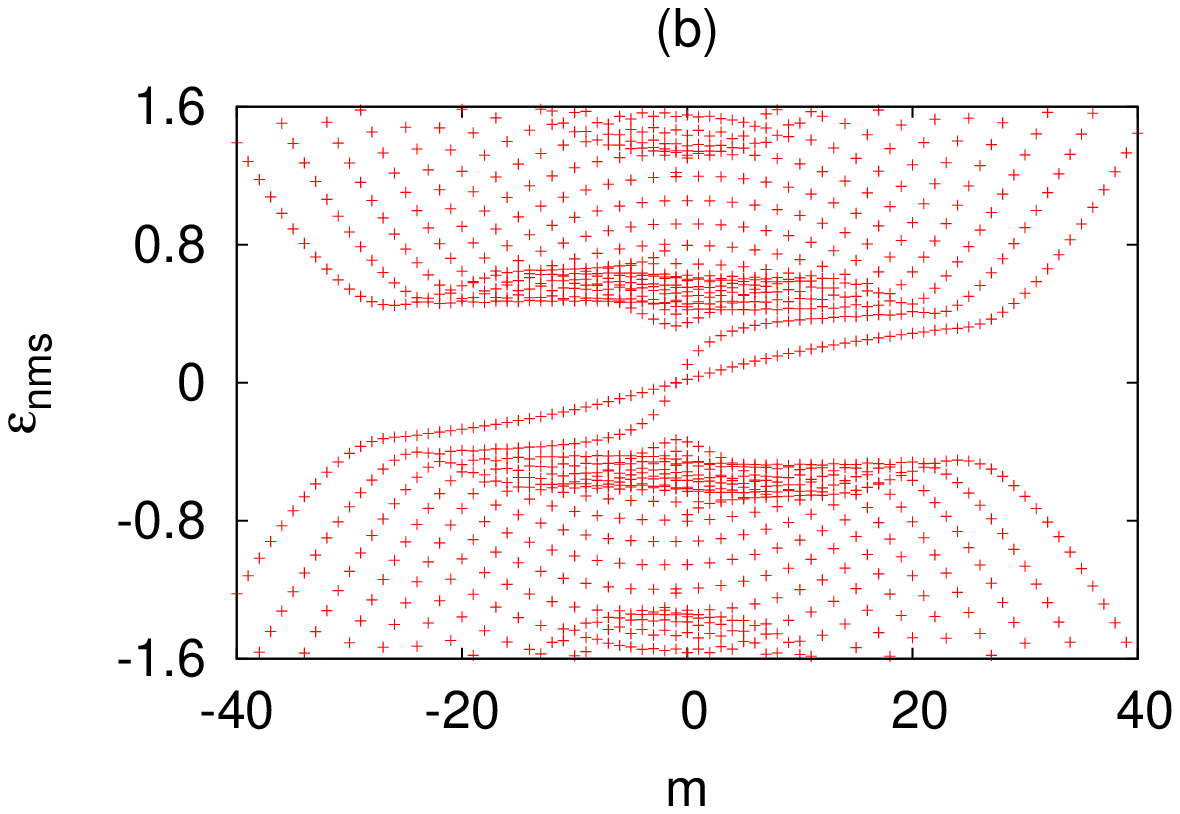}}}
\caption{\label{fig:spectrum} (Color online) 
The energy spectrum $\varepsilon_{nms}$ (in units of $\varepsilon_F$) is shown as a 
function of planar angular momentum $m$ when $s = 0$. Here, we set $1/(k_F a_s) = 0$ 
and $ \alpha = 0.5 k_F/M$, and vary $P$.  In (a) we choose $P = 0.2$ corresponding 
to a topologically-trivial gapped bulk SF phase, and in (b) we choose $P = 0.5$ 
corresponding to a topologically-nontrivial gapless bulk GSF phase. The linear branch
in (b) corresponds to the edge-bound states.
}
\end{figure}

In Fig.~\ref{fig:spectrum}, the spectra of energy eigenvalues $\varepsilon_{nms}$ are
shown as a function of planar angular momentum $m$ for the $s = 0$ sector. The 
spectrum rapidly becomes symmetric around $m = 0$ with increasing $s$, since 
the vortex core and edge states disappear when $|s| \sim 1$. 
Here, we choose $\alpha$ and $P$ such that Fig.~\ref{fig:spectrum}(a) corresponds to a 
topologically-trivial gapped bulk SF phase, and Fig.~\ref{fig:spectrum}(b) corresponds 
to a topologically-nontrivial gapless bulk GSF phase (see the thermodynamic phase 
diagram given in Fig.~\ref{fig:pd}). First of all, we note that the excitation spectra shown in
these figures have the necessary symmetry $\varepsilon_{nms} = - \varepsilon_{n,-(m+2),s}$, 
which follows from the particle-hole symmetry of the Hamiltonian. In addition, 
a second branch of continuum spectra appears in both cases and on both positive and 
negative energy regions when $|\varepsilon_{nms}| \gtrsim 1.2\varepsilon_F$. 
This is similar to what happens in a thermodynamic system, for which the excitation 
spectrum has two quasiparticle and two quasihole branches 
when $\alpha \ne 0$~\cite{subasi, carlos}.

When $\alpha = 0$, the positive- and negative-energy spectra are connected 
by a single branch of discrete Andreev-like bound states~\cite{caroli, gygi}. 
While the visible discreteness of the continuum spectrum is a finite-size effect 
and the spectrum becomes continuous only in the thermodynamic limit ($k_F R \to \infty$), the 
discreteness of the bound states is insensitive to the system size since 
these states are strongly localized around the vortex core. The lowest-energy 
quasiparticle excitation requires a minigap of order $\Delta_0^2/(2\varepsilon_F) \ll \Delta_0$.
When $\alpha \ne 0$, the energy spectrum of the SF phase shown in 
Fig.~\ref{fig:spectrum}(a) is very similar to the usual population-balanced 
$s$-wave superfluids, for which the bulk energy spectrum is also gapped. 
There are only a few discrete core-bound states appearing within the bulk 
energy gap in Fig.~\ref{fig:spectrum}(a), since the bulk order parameter 
$\Delta_0 \simeq 0.7 \varepsilon_F$ is quite large for $P = 0.2$ leading also to 
a large minigap. However, the bulk order parameter decreases to 
$\Delta_0 \simeq 0.5 \varepsilon_F$ when $P = 0.5$, leading to a smaller minigap
in comparison to $P = 0.2$ case, and hence a larger number of core-bound 
states as can be clearly seen in Fig.~\ref{fig:spectrum}(b).

In contrast, we see a major difference in the energy spectrum of the GSF phase
as shown in Fig.~\ref{fig:spectrum}(b). In addition to the branch of discrete core-bound
states that is also present in the SF phase, there is a second branch of bound states 
which are strongly localized around the edge of the cylinder. These states result 
from Andreev scattering at the rigid walls of the cylinder, and their spectrum is linear
in energy within the continuum gap~\cite{note}. We find that the lowest positive-energy 
and highest negative-energy bound states have $m = -1$ and $s = 0$, and their energies 
are $\varepsilon_0 \approx 4.33\times10^{-4} \varepsilon_F$  and 
$\varepsilon_{0'} \approx -4.33\times10^{-4} \varepsilon_F$, respectively. 
This is not a coincidence since we know that the energy spectrum has 
$\varepsilon_{nms} = - \varepsilon_{n,-(m+2),s}$ symmetry, and given that the spectrum is 
expected to have a two-fold degenerate zero-energy bound states, i.e. a pair of Majorana 
quasiparticles, in the thermodynamic limit, they must occur at $m = -1$. However, 
hybridization between the core- and edge-bound states (see below) lifts this degeneracy 
in a finite system, and the zero-energy bound states split in energy as we find here.

\begin{figure} [htb]
\centerline{\scalebox{0.6}{\includegraphics{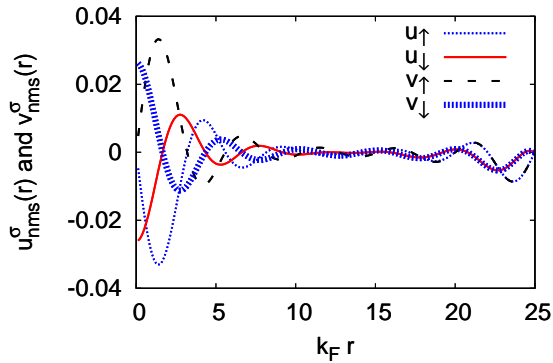}}}
\caption{\label{fig:wf} (Color online) 
The radial wave functions $u_{nms}^\sigma(r)$ and $v_{nms}^\sigma(r)$ (in units of $1/k_F$) 
are shown as a function of radial distance $r$ (in units of $1/k_F$) for the lowest 
positive-energy bound state. For the parameters considered in Fig.~\ref{fig:spectrum}(b), 
$\varepsilon_0 \approx 4.33\times10^{-4} \varepsilon_F$ and it occurs at $m = -1$ and $s = 0$.
}
\end{figure}

In Fig.~\ref{fig:wf}, the radial wave functions $u_{nms}^\sigma(r)$ and $v_{nms}^\sigma(r)$ 
of the lowest positive-energy bound state with energy $\varepsilon_0$ are shown as a 
function of radial distance $r$, for the parameters considered in Fig.~\ref{fig:spectrum}(b). 
(Due to particle-hole symmetry, the radial wave functions for the highest negative-energy 
bound state with energy $\varepsilon_{0'}$ can simply be obtained by changing 
$u_{nms}^\sigma(r) \rightarrow v_{nms}^\sigma(r)$ and vice versa.)
We note that while the wave functions have $u_{nms}^\sigma(r) = - v_{nms}^\sigma(r)$ 
symmetry near the vortex core, they have $u_{nms}^\sigma(r) = v_{nms}^\sigma(r)$ symmetry
near the edge. These are consistent with the symmetries of Majorana quasiparticles~\cite{mao},
for which the associated quasiparticle operators are self-Hermitian, i.e. a Majorana 
quasiparticle is its own anti-quasiparticle. This is clearly seen from the 
Bogoliubov-Valatin quasiparticle creation operator
$
\gamma_n^\dagger = \int d\mathbf{r} \sum_\sigma [ u_n^\sigma(\mathbf{r}) \psi_\sigma^\dagger(\mathbf{r})
+  v_n^\sigma(\mathbf{r}) \psi_\sigma(\mathbf{r})]
$
evaluated at $m = -1$ and $s = 0$. Since the Majorana quasiparticles always 
come in pairs, they appear simultaneously but away from each other in real space. 
In our single vortex line, while one of them is mostly localized at the vortex core, 
the other one is mostly localized at the edge, with some degree of hybridization 
between them due to finite-size effects. The hybridization is clearly seen in the 
wave functions shown in Fig.~\ref{fig:wf}. 
We note that due to this coupling between the Majorana core-
and edge-bound states, their two-fold $\varepsilon_{nms} = 0$ degeneracy is lifted, 
causing a small level splitting as discussed above. Increasing the separation between 
Majorana core- and edge-bound states, i.e. when $k_F R \to \infty$, weakens the 
hybridization such that both bound states eventually become degenerate in energy
with $\varepsilon_{nms} = 0$. When this happens, the core quasiparticle is well-localized 
around the vortex core with $u_{nms}^\sigma(r) = - v_{nms}^\sigma(r)$ symmetry, 
and the edge quasiparticle is well-localized around the edge with 
$u_{nms}^\sigma(r) = v_{nms}^\sigma(r)$ symmetry, without any hybridization between 
the two.

So far, we have established a major difference between the energy spectra of 
SF and GSF phases, which is mainly due to the appearance of edge- and Majorana
zero-energy bound states, and this difference leaves its signatures in various 
observables as discussed next.

\subsection{Single-particle density of states}
\label{sec:dos}

For instance, the local single-particle density of $\sigma$ states
$
D_\sigma(\mathbf{r}, \omega) = \sum_n [ |u_n^\sigma(\mathbf{r})|^2 \delta(\omega - \varepsilon_n)
+ |v_n^\sigma(\mathbf{r})|^2 \delta(\omega + \varepsilon_n)],
$
where $\delta(x)$ is the delta function, as well as the integrated single-particle 
density of $\sigma$ states
$
D_\sigma(\omega) = \int d\mathbf{r} D_\sigma(\mathbf{r},\omega)
$
provide direct evidences for the existence of edge- and Majorana zero-energy 
bound states as shown below. 
In particular, for a vortex line, and after using the orthonormality 
conditions for the Bessel functions, $D_\sigma(\omega)$ reduces to 
\begin{align}
D_\sigma(\omega) = \sum_{nmsj} [(c_{nmsj}^\sigma)^2  \delta(\omega - \varepsilon_{nms}) + (d_{nmsj}^\sigma)^2 \delta(\omega + \varepsilon_{nms} ].
\end{align}
We use a small spectral broadening ($0.01\varepsilon_F$) to regularize
the delta functions in our numerical calculations.

\begin{figure} [htb]
\centerline{\scalebox{0.6}{\includegraphics{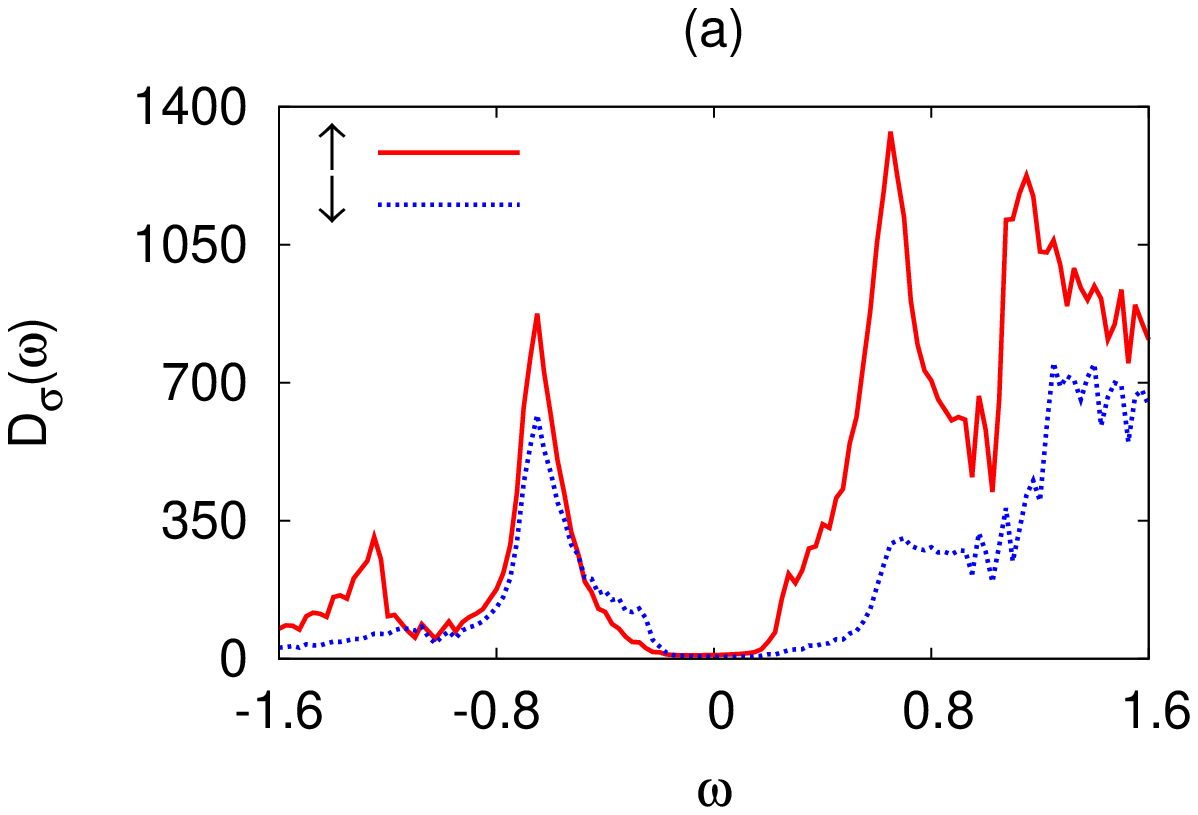}}}
\centerline{\scalebox{0.6}{\includegraphics{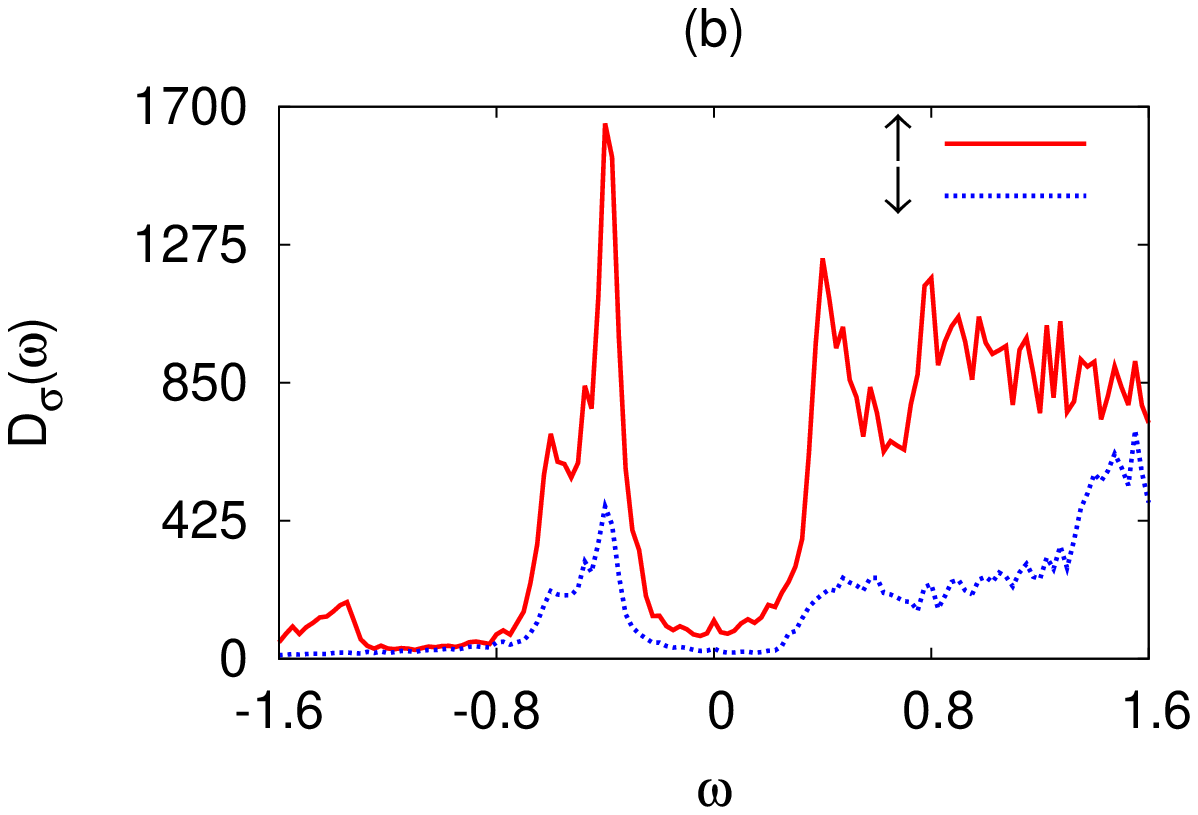}}}
\caption{\label{fig:dos} (Color online) 
The integrated density of $\sigma$ states $D_\sigma(\omega)$ (in units of $1/\varepsilon_F$) 
are shown as a function of energy $\omega$ (in units of $\varepsilon_F$), 
for the parameters considered in Fig.~\ref{fig:spectrum}. 
}
\end{figure}

In Fig.~\ref{fig:dos}, the integrated density of $\sigma$ states $D_\sigma(\omega)$ are shown 
as a function of energy $\omega$, for the parameters considered in Fig.~\ref{fig:spectrum}. 
We see that while the density of states vanishes around $\omega = 0$ in the SF phase, 
due to the presence of a gap in the energy spectrum, it is finite in the GSF phase with 
very small peaks around $\omega = 0$. These peaks are due to the presence of discrete 
core and edge states within the continuum gap in energy, and they are most clearly seen 
in the majority ($\uparrow$) component. We also note that the appearance of a second 
continuum branch in the excitation spectrum increases the density of states 
considerably when $|\varepsilon_{nms}| \gtrsim 1.2 \varepsilon_F$. 
Next, we analyze the local probability-current density of $\sigma$ fermions,
which also shows signatures for the edge- and Majorana zero-energy bound states.

\subsection{Probability-current density}
\label{sec:current}

Similar to the usual $\alpha = 0$ treatment, the quantum mechanical probability-current 
operator for $\sigma$ fermions can be identified from the continuity equation. 
While the presence of a spin-orbit coupling leads to additional terms in the total 
particle current operator, these terms do not contribute to the current since the 
expectation value $\langle \psi_\uparrow^\dagger(\mathbf{r}) \psi_\downarrow(\mathbf{r}) \rangle = 0$.
Therefore, using the Bogoliubov-Valatin transformations, the local current density 
$
\mathbf{J}_\sigma (\mathbf{r}) = [1/(2M_\sigma i)]  
\langle \psi_\sigma^\dagger(\mathbf{r}) \nabla \psi_\sigma(\mathbf{r}) - H.c. \rangle
$
circulating around a single vortex line becomes
$
\mathbf{J}_\sigma (\mathbf{r}) = [1/(2M_\sigma i)] \sum_n
[{u_n^\sigma}^*(\mathbf{r}) \nabla u_n^\sigma(\mathbf{r}) f(\varepsilon_n)
+ {v_n^\sigma}^*(\mathbf{r}) \nabla v_n^\sigma(\mathbf{r}) f(-\varepsilon_n) - H.c.],
$
where $H.c.$ is the Hermitian conjugate.
Since $\mathbf{J}_\sigma(\mathbf{r})$ circulates along the $\mathbf{\widehat{\theta}}$ 
direction, i.e. $ \mathbf{J}_\sigma(\mathbf{r}) = J_\sigma(r) \mathbf{\widehat{\theta}}$, we find
\begin{align}
\label{eqn:curup}
J_\uparrow(r) &= \frac{1}{2\pi M_\uparrow r} \sum_{nms} \big\lbrace 
m[\sum_j c_{nmsj}^\uparrow \phi_{j,m}(r)]^2 f(\varepsilon_{nms}) \nonumber \\
&- (m+2) [\sum_j d_{nmsj}^\uparrow \phi_{j,m+2}(r)]^2 f(-\varepsilon_{nms}) \big\rbrace, \\
\label{eqn:curdo}
J_\downarrow(r) &= \frac{1}{2\pi M_\downarrow r} \sum_{nms} \big\lbrace 
(m+1)[\sum_j c_{nmsj}^\downarrow \phi_{j,m+1}(r)]^2 f(\varepsilon_{nms}) \nonumber \\
&- (m+1) [\sum_j d_{nmsj}^\downarrow \phi_{j,m+1}(r)]^2 f(-\varepsilon_{nms}) \big\rbrace,
\end{align}
for the strengths of the particle-current densities.

\begin{figure} [htb]
\centerline{\scalebox{0.6}{\includegraphics{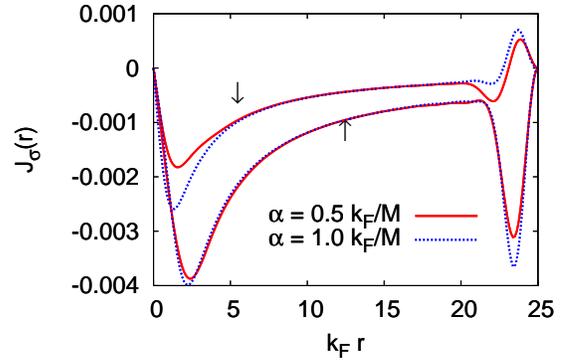}}}
\caption{\label{fig:current} (Color online) 
The probability current density of $\sigma$ fermions (in units of $M/k_F^3$) are shown as 
a function of radial distance $r$ (in units of $1/k_F$),
for the parameters considered in Fig.~\ref{fig:density}.
}
\end{figure}

In Fig.~\ref{fig:current}, the probability-current density of $\sigma$ fermions
are shown as a function of radial distance $r$, for the parameters considered in Fig.~\ref{fig:density}.
When $\alpha = 0$, the core-bound states have negative (diamagnetic) and the 
continuum states have positive (paramagnetic) contribution to $J_\sigma(r)$.
This leads to a nonmonotonic $J_\sigma(r)$ which first increases as $\propto r$ 
and then decreases as $\propto 1/r$~\cite{gygi}. The latter behavior is due to
the saturation of the superfluid density for long distances away from the vortex core. 
Therefore, a maximum peak current occurs at some distance away from the vortex core. 
When $\alpha \ne 0$, the major difference is at the edge. The $\uparrow$ and $\downarrow$ 
currents flow in opposite directions, and their magnitude increases with increasing $\alpha$.
Such a counterflow of mass currents occurs even for the SF phase (not shown).
Since the sums in Eqs.~(\ref{eqn:curup}) and~(\ref{eqn:curdo}) are over states 
with $\varepsilon_{nms} \ge 0$, counterflowing edge currents result from the asymmetry of 
the energy spectrum around $m = 0$, due to the presence of edge states.

\section{Conclusions}
\label{sec:conc}

To conclude, we considered a Rashba-type spin-orbit coupled Fermi gas, and 
used the BdG formalism to analyze a single vortex line along a finite cylinder with
a periodic boundary condition. 
When the populations of the pseudo-spin components are sufficiently imbalanced,
depending on the strength of the spin-orbit coupling, we showed that core- and 
edge-bound states as well as Majorana zero-energy quasiparticles appear in the energy 
spectrum. These states leave signatures in the density of single-particle states and 
particle-current density, and particularly, we found that the 
pseudo-spin components counterflow near the edge of the cylinder, the strength of 
which increases with increasing spin-orbit coupling.

While preparing this work, we became aware of a closely related work~\cite{liu}, 
where the vortex core and edge states are
analyzed for a trapped two-dimensional Fermi gas. For the most parts, our work
is consistent with their findings. However, in contrast to our finite-cylinder setup, 
where the system is either in an SF or a GSF phase, depending on the parameter 
regime, both SF and GSF phases may also coexist in a trap in different regions. The 
possibility of such a phase coexistence again leads to Andreev scattering at the 
SF-GSF phase boundary, giving rise to an additional branch of edge-bound states.

\section{Acknowledgments}
This work is supported by the Marie Curie International Reintegration 
(Grant No. FP7-PEOPLE-IRG-2010-268239), Scientific and Technological 
Research Council of Turkey (Career Grant No. T\"{U}B$\dot{\mathrm{I}}$TAK-3501-110T839), 
and the Turkish Academy of Sciences (T\"{U}BA-GEB$\dot{\mathrm{I}}$P).


\begin{thebibliography}{99}

\bibitem{nist1} Y.-J. Lin, R. L. Compton, A. R. Perry, W. D. Phillips, J. V. Porto, and I. B. Spielman, Phys. Rev. Lett. \textbf{102}, 130401 (2009).
\bibitem{nist2} Y.-J. Lin,     Y.-J. Lin, K. Jim\'enez-Garc\'ia, and I. B. Spielman, Nature (London) \textbf{471}, 83 (2011).

\bibitem{gong} M. Gong, S. Tewari, and C. Zhang, Phys. Rev. Lett. \textbf{107}, 195303(2011);
G. Chen, M. Gong, and C. Zhang, Phys. Rev. A \textbf{85}, 013601 (2012).
\bibitem{shenoy} J. P. Vyasanakere, S. Zhang, and V. B. Shenoy, Phys. Rev. B \textbf{84}, 014512 (2011).
\bibitem{zhai} Z. Q. Yu and H. Zhai, Phys. Rev. Lett. \textbf{107}, 195305 (2011); 
H. Zhai, arXiv:1110.6798 (2011).
\bibitem{hui} Hui Hu, L. Jiang, X.-J. Liu, and Han Pu, Phys. Rev. Lett. \textbf{107}, 195304 (2011); 
Phys. Rev. A \textbf{84}, 063618 (2011).

\bibitem{subasi} M. Iskin and A. L. Suba{\c s}{\i}, Phys. Rev. Lett. \textbf{107}, 050402 (2011);
Phys. Rev. A \textbf{84}, 043621 (2011).
\bibitem{wyi} W. Yi and G.-C. Guo, Phys. Rev. A \textbf{84}, 031608(R) (2011).
\bibitem{carlos} K. Seo, Li Han, and C. A. R. S\'a de Melo, arXiv:1108.4068 and arXiv:1110.6364 (2011).

\bibitem{he} L. He and X. G. Huang, arXiv: 1109.5577 (2011).
\bibitem{ghosh} S. K. Ghosh, J. P. Vyasanakere, and V. B. Shenoy, Phys. Rev. A, \textbf{84}, 053629 (2011).
\bibitem{yang} B. Huang and S. Wan, arXiv:1109.3970 (2011); X. Yang and S. Wan, arXiv:1111.4277 (2011).
\bibitem{zhou} J. Zhou, W. Zhang, and W. Yi, Phys. Rev. A \textbf{84}, 063603 (2011).
\bibitem{duan} J. N. Zhang. Y. H. Chan, and L. M. Duan, arXiv: 1110.2241 (2011).
\bibitem{zhang} K. Zhou and Z. Zhang, Phys. Rev. Lett. \textbf{108}, 025301 (2012).
\bibitem{liao} R. Liao, Y. Y. Xiang, and W.-M. Liu, arXiv:1110.5818 (2011).

\bibitem{moore} G. Moore and N. Read, Nucl. Phys. B \textbf{360}, 362 (1991).

\bibitem{volovik} G. Volovik, JETP Lett. \textbf{70}, 609 (1999).
\bibitem{read} N. Read and D. Green, Phys. Rev. B 61, 10267 (2000).

\bibitem{tewari} S. Tewari, S. Das Sarma, and D.-H. Lee, Phys. Rev. Lett. \textbf{99}, 037001 (2007).
\bibitem{mizushima} T. Mizushima, M. Ichioka, and K. Machida, Phys. Rev. Lett. \textbf{101}, 150409 (2008).

\bibitem{fu} L. Fu and C. L. Kane, Phys. Rev. Lett. \textbf{100}, 096407 (2008).

\bibitem{oreg} Y. Oreg, G. Refael, and F. von Oppen, Phys. Rev. Lett. \textbf{105}, 177002 (2010).
\bibitem{alicea} J. Alicea, Y. Oreg, G. Refael, F. von Oppen, and M. P. A. Fisher, Nature Phys \textbf{7}, 412 (2011).

\bibitem{sau} J. D. Sau, R. M. Lutchyn, S. Tewari, and S. Das Sarma, Phys. Rev. Lett. \textbf{104}, 040502 (2010).
\bibitem{mao} Li Mao and C. Zhang, Phys. Rev. B \textbf{82}, 174506 (2010).

\bibitem{gygi}  F. Gygi and M. Schl\"uter, Phys. Rev. B \textbf{43}, 7609 (1991).

\bibitem{takahashi} M. Takahashi, T. Mizushima, M. Ichioka, and K. Machida, Phys. Rev. Lett. \textbf{97}, 180407 (2006).

\bibitem{caroli} C. Caroli, P. de Gennes, and J. Matricon, Phys. Lett. \textbf{9}, 307 (1964).

\bibitem{note} This is similar to what happens for a single vortex in two-dimensional chiral 
$p$-wave superfluids~\cite{mizushima}, where Majorana core- and edge-bound states 
also appear in the weakly-interacting BCS regime.

\bibitem{liu} X.-J. Liu, L. Jiang, Han Pu, and Hui Hu, arXiv:1111.1798 (2011). 
Our formalism and theirs have minor differences, e.g. $v_n^\sigma(\mathbf{r})$ amplitudes 
differ by a minus sign in their Bogoliubov-Valatin transformation, leading to some 
other presentational differences.

\end{thebibliography}
\end{document}